\documentclass[english,prd,superscriptaddress,nofootinbib,preprintnumbers,twocolumn,showpacs]{revtex4-1}
\usepackage[english]{babel}
\usepackage{amsmath}
\usepackage{amsfonts}
\usepackage{amssymb}
\usepackage{epsfig}
\usepackage{graphics,psfrag,rotating}
\usepackage{graphicx}
\usepackage[colorlinks=true,
            linkcolor=blue,
          urlcolor=gray,
            citecolor=red]{hyperref}

\newcommand{\nn}{\nonumber\\}

\def\3nab{\tilde{\nabla}}

\def\be {\begin{equation}}
\def\ee {\end{equation}}
\def\ba {\begin{align}}
\def\ea {\end{align}}

\def\bc {\begin{center}}
\def\ec {\end{center}}
\def\case#1/#2{\frac{#1}{#2}}

\newcommand{\bver}{\begin{verbatim}}
\newcommand{\ever}{\end{verbatim}}

\newcommand{\bea}{\begin{eqnarray}}
\newcommand{\eea}{\end{eqnarray}}
\newcommand{\beaa}{\begin{eqnarray*}}
\newcommand{\eeaa}{\end{eqnarray*}}

\newcommand{\e}{\mathrm{e}}

\def\case#1/#2{\textstyle\frac{#1}{#2}}

\begin{document}
%%%%%%%%%%%%%%%%%%%%%%%%%%%%%%%%%%%%%%%

\title{Analyzing modified unimodular gravity via Lagrange multipliers}

\author{Diego S\'aez-G\'omez\footnote{dsgomez [at] fc.ul.pt}}
\affiliation{Departamento de F\'isica \& Instituto de Astrof\'isica e Ci\^encias do Espa\c{c}o,
Faculdade de Ci\^encias da Universidade de Lisboa, Edif\'icio C8, Campo Grande, P-1749-016
Lisbon, Portugal}

\pacs{04.50.Kd, 98.80.-k, 98.80.Cq, 12.60.-i}

%%%%%%%%%%%%%%%%%%%%%%%%%%%%%%%%%%%%%
\begin{abstract} 
The so-called unimodular version of General Relativity is revisited. Unimodular gravity is constructed by fixing the determinant of the metric, what leads to the trace-free part of the equations instead of the usual Einstein field equations. Then, a cosmological constant   naturally arises as an integration constant. While unimodular gravity turns out equivalent to General Relativity (GR) at the classical level, it provides important differences at the quantum level. Here we extend the unimodular constraint to some extensions of General Relativity that have drawn a lot of attention over the last years, as $f(R)$ gravity (or its scalar-tensor picture) and Gauss-Bonnet gravity. The corresponding unimodular version of such theories is constructed as well as the conformal transformation that relates the Einstein and Jordan frames for these non-minimally coupled theories. From the classical point of view, the unimodular versions of such extensions are completely equivalent to their originals, but an effective cosmological constant arises naturally, what may provide a richer description of the universe evolution. Here we analyze the case of Starobisnky inflation and compared with the original one.

\end{abstract} 
%%%%%%%%%%%%%%%%%%%%%%%%%%%%%%%%%%%%%
%\date{\today}
\maketitle

%%%%%%%%%%%%%%%%%%%%%%%%%%%%%%%%%%%%%
\section{Introduction}
%%%%%%%%%%%%%%%%%%%%%%%%%%%%%%%%%%%%%

Over the last decades, the so-called {\it cosmological constant problem} has been one of the major challenges in theoretical physics. The issue refers to the absence of gravitational effects, particularly at the cosmological level, of the vacuum energy density predicted by quantum field theories, or better said, the impossibility of fine tuning properly its counterterms, what is known as radiative instability (for a review see Refs.~\cite{Weinberg:1988cp} and \cite{Padilla:2015aaa}). On the other hand, after the discovery of some deviations of the luminosity distances of Supernovae Ia in 1998, what was then interpreted as a consequence of the acceleration of the universe expansion (and confirmed by other proofs later), the best and most accepted model that can explain such behaviour lies on the presence of a cosmological constant in the gravitational field equations, which in principle should be connected somehow to the vacuum energy density. However, the required cosmological constant for the acceleration of the expansion (of the order the Hubble parameter today) is around 120 orders of magnitude smaller than the one predicted by quantum field theories. Hence, here the problem arises, how to drop down the huge value of vacuum fluctuations. In this sense, there have been plenty of proposals, which include  a possible symmetry that protects the cosmological constant in the same sense as chiral symmetry does with the electron mass as well as supersymmetry attempts. In addition, an alternative way, which may include the dark energy models/modified gravities, tries to suppress such large value by additional fields or modifications of General Relativity (GR). In this sense, there have been  plenty of dark energy models proposed, which may play that role, see \cite{dark_energy} and \cite{Nojiri:2010wj}. However, rather than solving the problem, the former always requires a precise fine tuning as well (Weinberg's no go theorem).\\

An alternative widely studied in the literature is the so-called unimodular gravity (see Refs.~\cite{Weinberg:1988cp,Padilla:2015aaa,Finkelstein:2000pg,Alvarez:2006uu,Alvarez:2005iy,Alvarez:2007nn,Alvarez:2008zw,Alvarez:2009ga,Buchmuller:1988wx,Henneaux:1989zc,Padilla:2014yea}). The theory fixes the determinant of the metric, such that the field equations are given by the trace-free part of GR's field equations. From the classical point of view, fixing the determinant of the metric provides a cosmological constant that naturally arises as an integration constant after applying the corresponding geometrical  identities, what at the cosmological level may be a way of understanding the problem of dark energy \cite{Finkelstein:2000pg}. The unimodular constraint can be implemented in several ways, all of them leading to the same classical theory, as shown in the literature \cite{Finkelstein:2000pg,Alvarez:2006uu,Alvarez:2005iy,Alvarez:2007nn,Alvarez:2008zw,Alvarez:2009ga,Buchmuller:1988wx,Henneaux:1989zc,Padilla:2014yea}. However, in spite of the theory is equivalent to GR at the classical level, the equivalence is not clear at the quantum one, where a great effort has been done in order to get a better understanding of the features of the theory \cite{Alvarez:2005iy,Padilla:2014yea}. When the theory is analyzed in quantum mechanics, radiative instability is absence for this effective cosmological constant, being one of the most interesting features of the theory, since it can suppress the large contribution of the vacuum energy density \cite{Alvarez:2007nn}. The absence of radiative instability has been shown in the literature by using different approaches, for instance by the existence of a shift symmetry in the classical field equations that remove the contributions from the quantum vacuum, but also by the evaluation of the renormalization group equation for the cosmological constant. In addition, unimodular gravity may be distinguished from GR by some observables, as it may lead to a different concept of mass \cite{Alvarez:2009ga}. Hence, unimodular gravity may provide not only a way of understanding the cosmological constant problem better but also to shed some light to the dark energy issue.   \\
%, while the cosmological perturbations may be affected, the differences are suppressed for adiabatic perturbations (see Ref.~\cite{Gao:2014nia}).\\

In this sense, an extension of unimodular gravity has been recently proposed, where more general actions rather than the Hilbert-Einstein action are considered \cite{Eichhorn:2015bna,Nojiri:2015sfd}. Note that modified gravities as $f(R)$ gravity, have drawn a lot of attention over the last years, as they can realize the cosmological history, being alternatives to dark energy/inflaton. In addition, some of them are able to satisfy the last observational constraints with great accuracy, such as Starobinsky inflation \cite{staro}, or the so-called Hu-Sawicki model for late-time acceleration \cite{Hu:2007nk}. Hence, the analysis of such extensions within the unimodular-like framework may provide interesting features. \\

With this aim, this paper is devoted to the analysis of some generalizations of unimodular gravity at the classical level. Firstly, we carefully reconstruct such extensions departing from variational principles by using a Lagrange multiplier which imposes the unimodular constraint, leading to the trace-free part of the field equations. We show that as in the case of unmodular gravity, any extension leads to the same result, i.e. a cosmological constant arises naturally in the field equations, recovering full diffeomorphisms. We also analyze how conformal transformations affect the gauge choice imposed initially, and the effects of unimodular gravity in the Einstein frame. Finally, some cosmological solutions are obtained for $f(R)$ gravity and Gauss-Bonnet gravities, while Starobinsky inflation is also analyzed, where we find a constraint on the merging constant in order to keep Starobinsky predictions.\\

The paper is organized as follows: Section \ref{Section_General} gives a brief review on unimodular gravity. In section \ref{General_unimodular_grav}, we introduce $f(R)$ and Gauss-Bonnet unimodular gravity.  In Section \ref{Einstein_sect}, the conformal transformation is analyzed and the corresponding unimodular version is obtained in the Einstein frame. Then, Section \ref{cosmo_sect} is devoted to the analysis of cosmological solutions. Finally, section \ref{Conclusions_sect} gathers the conclusions.\\

%%%%%%%%%%%%%%%%%%%%%%%%%%%%%%%%%%%%%
\section{Unimodular gravity}
\label{Section_General}
%%%%%%%%%%%%%%%%%%%%%%%%%%%%%%%%%%%%%

Unimodular gravity is constructed in such a way that the determinant of spacetime metric is not dynamical but is restricted to be:
\be
\sqrt{-g}=s_0\ ,
\label{1.10}
\ee
which fixes the determinant of the metric to be a constant $s_0$. As states in \cite{Weinberg:1988cp}, {\it just because we use a generally covariant formalism does not mean that we are committed to treating all components of the metric as dynamical fields.} Then, restricted variations of the action with respect to the metric have to be null only for those which keep the determinant fixed,
\be
g_{\mu\nu}\delta g^{\mu\nu}=0\ ,
\label{1bis}
\ee
The variation of the metric can be then written in terms of the unconstrained variation as:
\be
\delta g^{\mu\nu}=\delta_{u} g^{\mu\nu}-\frac{1}{4}g^{\mu\nu}g_{\lambda\gamma}\delta_{u} g^{\lambda\gamma}\ .
\label{2bis}
\ee
The gravitational field equations are obtained by varying the gravitational action $S_G$, which can be also expressed in terms of the unconstrained variation, leading to 
\be
\frac{\delta S_G}{\delta g^{\mu\nu}}=\frac{\delta S_G}{\delta_{u} g^{\mu\nu}}-\frac{1}{4}g_{\mu\nu}g^{\lambda\gamma}\frac{\delta S_G}{\delta_{u} g^{\lambda\gamma}}\ .
\label{3bis}
\ee
These are precisely the traceless part of the gravitational field equations, which for the Hilbert-Einstein action leads to the traceless part of the Einstein's field equations:
\be
R_{\mu\nu}-\frac{1}{4}g_{\mu\nu}R=\kappa^2\left(T_{\mu\nu}-\frac{1}{4}g_{\mu\nu}T\right)\ ,
\label{1.1}
\ee
where $R_{\mu\nu}$ is the usual Ricci tensor and $R$ its trace, while $T_{\mu\nu}=\frac{2}{\sqrt{-g}}\frac{\partial S_m}{\partial g_{\mu\nu}}$ is the matter energy-momentum tensor, and $\kappa^2=8\pi G$. Contrary to General Relativity, the field equations (\ref{1.1}) are not divergence free, 
\be
\nabla_{\mu}\left(R_{\mu\nu}-\frac{1}{4}g_{\mu\nu}R-\kappa^2T_{\mu\nu}-\frac{1}{4}g_{\mu\nu}T\right)=0\ .
\label{1.2}
\ee
Then, by using the Bianchi identities, which holds
\be
\nabla_{\mu}\left(R_{\mu\nu}-\frac{1}{2}g_{\mu\nu}R\right)=0\ ,
\label{1.3}
\ee
and the energy conservation,
\be
\nabla_{\mu}T_{\mu\nu}=0\ ,
\label{1.4}
\ee
The divergence of the field equations (\ref{1.2}) yields
\be
\nabla_{\mu}\left(R+\kappa^2 T\right)=0\ ,
\label{1.5}
\ee
which is the so-called integrability condition that after integrating leads to
\be
R+\kappa^2T=4\lambda_0=\text{constant}\ ,
\label{1.6}
\ee
where $\lambda_0$ is an integration constant. Hence, by inserting (\ref{1.6}) in the field equations (\ref{1.1}), we get
\be
R_{\mu\nu}-\frac{1}{2}g_{\mu\nu}R+g_{\mu\nu}\lambda_0=\kappa^2T_{\mu\nu}\ ,
\label{1.7}
\ee
where the usual Einstein field equations are recovered, being $\lambda_0$ a cosmological constant. This is the great success of unimodular gravity, since a cosmological constant emerges naturally as an integration constant by departing from the trace-free part of the Einstein equations. Such constant may compensate the large value of the vacuum energy density. In addition, since the integrability condition (\ref{1.5}) recovers the usual General Relativity equations, any prediction from the former turns out a prediction of unimodular gravity, what avoids any possible discrepancy with well tested experiments. \\

Alternatively, unimodular gravity can be implemented through a variational principle with unrestricted variations of the metric by assuming transverse diffeomorphisms (TDiff) instead of the full diffeomorphisms \cite{Alvarez:2006uu,Alvarez:2005iy,Alvarez:2007nn}, which gives rise to the appearance of a scalar field that represents the determinant of the metric. Such extra degree of freedom can be removed by an additional Weyl symmetry (WTDiff), \cite{Alvarez:2006uu,Alvarez:2005iy,Alvarez:2008zw}. Moreover, unimodular gravity can be also obtained by using a Lagrange multiplier in the action as follows \cite{Buchmuller:1988wx},
\be
S=\frac{1}{2\kappa^2}\int dx^4 \left[\sqrt{-g}R-2\lambda(\sqrt{-g}-s_0)\right]+S_m\ ,
\label{1.8}
\ee
where $s_0$ is a constant and $\lambda$ is the Lagrange multiplier, which in principle is dynamical. Note that the last term in (\ref{1.8}) breaks the full diffeomorphisms invariance, since it fixes the determinant of the metric, restricting the group of symmetries. Then, by varying the action with respect to the metric, the field equations yield,
\be
R_{\mu\nu}-\frac{1}{2}g_{\mu\nu}R+g_{\mu\nu}\lambda=\kappa^2T_{\mu\nu}\ ,
\label{1.9}
\ee
while the variation with respect to $\lambda$ leads to the unimodular restriction (\ref{1.10}). Taking the trace of the field equations (\ref{1.9}),
\be
R+\kappa^2T=4\lambda(x)\ ,
\label{1.11}
\ee
This looks as (\ref{1.6}) except that in principle $\lambda=\lambda(x)$ is not a constant. Nevertheless, taking the divergence of equations (\ref{1.9}) together with the energy conservation $\nabla_{\mu}T_{\mu\nu}=0$, yields,
\be
\nabla_{\mu}\lambda=0\ \rightarrow \lambda=\lambda_0\ .
\label{1.12}
\ee
Then, the trace-free part of the equations follow and the previous result (\ref{1.7}) is obtained, in this case by means of the action (\ref{1.8}). Moreover, equivalently to (\ref{1.8}) one can depart from the Henneaux-Teitelboim action, leading to the same result \cite{Henneaux:1989zc}. Note that while all these implementations of unimodular gravity are classically equivalent, they are not at the quantum level (see \cite{Alvarez:2005iy}). However, as we focus just on classical aspects along the paper, we are assuming the action (\ref{1.8}) as the departing point for convenience, as shown below. 
%This is an important point, since most of the modifications of GR depart from generalizations of the gravitational action, i.e., from considering more complex actions than the Einstein-Hilbert one. Hence, extensions of unimodular gravity can be studied by means of a variational principle. In the next section, we study some generalizations.
%
%%%%%%%%%%%%%%%%%%%%%%%%%%%%%%%%%%%%%%%%%
\section{Generalizations of unimodular gravity}
\label{General_unimodular_grav}
%%%%%%%%%%%%%%%%%%%%%%%%%%%%%%%%%%%%%%%%%
%
Over the last years, some modifications of the Hilbert-Einstein action have been considered, particularly as infrared corrections to GR in order to provide a natural explanation to the late-time acceleration of the expansion \cite{Nojiri:2010wj}. Moreover, such modifications have been widely applied to inflation, where nowadays data seem to favor some of such models  . Within modified gravities, the so-called $f(R)$ gravity has drawn a lot of attention, whose principle states on a gravitational action given precisely by,
\be
S=\frac{1}{2\kappa^2}\int dx^4\sqrt{-g}\ f(R)+S_m\ ,
\label{2.1}
\ee
whose field equations are obtained by varying the action with respect to the metric, leading to
\be
R_{\mu\nu}f_R-\frac{1}{2}g_{\mu\nu}f+\left(g_{\mu\nu}\Box-\nabla_{\mu}\nabla_{\nu}\right)f_R=\kappa^2T_{\mu\nu}\ ,
\label{2.2}
\ee
where $f_R=\frac{\partial f}{\partial R}$. Generalization of unimodular gravity turns out now clear. As pointed in \cite{Nojiri:2015sfd}, the action (\ref{2.1}) has an unimodular $f(R)$ version which is constructed by fixing the determinant to be a constant, 
\be
S=\frac{1}{2\kappa^2}\int dx^4\left[\sqrt{-g}\ f(R)-2\lambda(\sqrt{-g}-s_0)\right]+S_m\ .
\label{2.3}
\ee
The field equations are then given by
\be
R_{\mu\nu}f_R-\frac{1}{2}g_{\mu\nu}f+\left(g_{\mu\nu}\Box-\nabla_{\mu}\nabla_{\nu}\right)f_R+g_{\mu\nu}\lambda=\kappa^2T_{\mu\nu}\ ,
\label{2.4}
\ee
While the variation of the action with respect to $\lambda$ leads to $\sqrt{-g}=s_0$. 
%The trace of (\ref{2.4}) is given by
%\be
%Rf_R-2f+3\Box f_R+4\lambda=\kappa^2T\ .
%\label{2.5}
%\ee
As in the previous section, taking the divergence of the field equations (\ref{2.4}) yields
\be
\nabla_{\mu}\lambda=0\ \rightarrow \lambda=\lambda_0\ ,
\label{2.6}
\ee
where we have used the identities $\nabla_{\mu}\left(R^{\mu\nu}-\frac{1}{2}Rg^{\mu\nu}\right)=0$ and $(\nabla_{\nu}\Box-\Box\nabla_{\nu})f_R=R_{\mu\nu}\nabla^{\mu}f_R$. Then, by using the trace of the field equations (\ref{2.4}), the following condition is provided
\be
-Rf_R+2f-3\Box f_R+\kappa^2T=4\lambda_0\ ,
\label{2.7}
\ee
which is the generalization of the integrability condition (\ref{1.6}). Hence, the usual $f(R)$ equations are recovered with an additional cosmological constant:
\be
R_{\mu\nu}f_R-\frac{1}{2}g_{\mu\nu}f+\left(g_{\mu\nu}\Box-\nabla_{\mu}\nabla_{\nu}\right)f_R+g_{\mu\nu}\lambda_0=\kappa^2T_{\mu\nu}\ ,
\label{2.8}
\ee
Equivalently, one may proceed to obtain the same result by starting from the trace-free part of (\ref{2.2}) as the field equations:
\begin{widetext}
\be
R_{\mu\nu}f_R-\frac{1}{2}g_{\mu\nu}f+\left(g_{\mu\nu}\Box-\nabla_{\mu}\nabla_{\nu}\right)f_R-\frac{1}{4}\left(Rf_R-2f+3\Box f_R\right)g_{\mu\nu}=\kappa^2\left(T_{\mu\nu}-\frac{1}{4}g_{\mu\nu}T\right)\ .
\label{2.9}
\ee
\end{widetext}
By using  $\nabla_{\mu}T^{\mu\nu}=0$ and the Bianchi identities, the divergence of (\ref{2.9}) yields
\be
\nabla_{\mu}\left(Rf_R-2f+3\Box f_R-\kappa^2T\right)=0\ ,
\label{2.10}
\ee
which is equivalent to (\ref{2.7}) after integrating and  the field equations (\ref{2.8}) are recovered. \\

Hence, it is straightforward to construct other generalizations of unimodular gravity by following the procedure described above. For instance, we may consider the so-called modified Gauss-Bonnet gravity,
\be
S=\frac{1}{2\kappa^2}\int dx^4\left[\sqrt{-g}\ (R+f(G)-2\lambda(\sqrt{-g}-s_0)\right]+S_m\ ,
\label{2.11}
\ee
where $G=R_{\mu\nu\lambda\sigma}R^{\mu\nu\lambda\sigma}-4R_{\mu\nu}R^{\mu\nu}+R^2$ is the Gauss-Bonnet topological invariant. The field equations are obtained by varying the action (\ref{2.11}) with respect to the metric \cite{Elizalde:2010jx},
\begin{widetext}
\begin{eqnarray}
R_{\mu\nu}-\frac{1}{2}g_{\mu\nu}R-\frac{1}{2}g_{\mu\nu}f(G)+2f_{G}RR_{\mu\nu}-4f_{G}R_{\mu\rho}R_{\nu}^{\;\;\rho}+2f_{G}R^{\mu\rho\sigma\tau}R_{\nu\rho\sigma\tau}\nonumber\\
+4f_{G}R_{\mu\rho\sigma\nu}R^{\rho\sigma}-2R\nabla_{\mu}\nabla_{\nu}f_{G}
+2g_{\mu\nu}R\Box f_{G}+4R_{\nu\rho}\nabla^{\rho}\nabla_{\mu}f_{G}+4R_{\mu\rho}\nabla^{\rho}\nabla_{\nu}f_{G} \nonumber\\ 
-4R_{\mu\nu}\Box f_{G}-4g_{\mu\nu}R^{\rho\sigma}\nabla_{\rho}\nabla_{\sigma}f_{G}+4R_{\mu\rho\nu\sigma}\nabla^{\rho}\nabla^{\sigma}f_{G}+\lambda g_{\mu\nu}=\kappa^2T^{\mu\nu}\ .
\label{2.12}
\end{eqnarray}
\end{widetext}
As above, by taking the divergence of the field equations, the condition $\nabla_{\mu}\lambda=0$ arises again, what leads to the integrability condition for this case:\\
\begin{widetext}
\bea
R+2f-2f_GR^2+4f_GR_{\mu\nu}R^{\mu\nu}-2f_GR^{\mu\nu\lambda\sigma}R_{\mu\nu\lambda\sigma}-4f_GR^{\mu\nu\lambda}_{\;\;\;\;\;\;\mu}R_{\nu\lambda}\\ \nonumber
-2R\Box f_G+8R^{\mu\nu}\nabla_{\mu}\nabla_{\nu}f_G+4R^{\mu\nu\lambda}_{\;\;\;\;\;\;\mu}\nabla_{\nu}\nabla_{\lambda}f_G+T=4\lambda_0\ .
\label{2.13}
\eea
\end{widetext}
Then, the usual modified Gauss-Bonnet gravity with an additional cosmological constant. Note that the same result is obtained when departing from the trace-free part of the field equations fro Gauss-Bonnet gravity, as was shown above for the case of $f(R)$ gravity. Hence, following any of the above procedures, unimodular gravity can be easily extended to other more complex actions. The result basically adds a cosmological constant to the field equations, equivalently to the case of Hilbert-Einstein unimodular gravity. \\
Alternatively to the Lagrange multiplier, one may depart from restricting variations over the gravitational action (\ref{3bis}), leading to the traceless part of the corresponding $f(R)$ or $f(R,G)$ action, as above. Other implementations of unimodular gravity can be also applied for these cases equivalently at the classical level. However, by using a Lagrange multiplier instead of other implementations of the unimodular condition, calculations are simplified when dealing with theories with higher order derivatives. In the next section, we analyze unimodular scalar-tensor theories (equivalent to $f(R)$ gravities) and its transformation to the so-called Einstein frame when applying a conformal transformation, which becomes also simpler when forcing the unimodular constraint by a Lagrange multiplier than other alternative -classically- equivalent implementations.

%%%%%%%%%%%%%%%%%%%%%%%%%%%%%%%%%%%%%%%%%%%%%%%%%%%%%%
\section{Conformal frames}
\label{Einstein_sect}
%%%%%%%%%%%%%%%%%%%%%%%%%%%%%%%%%%%%%%%%%%%%%%%%%%%%%%

As well known, $f(R)$ gravities can be expressed in terms of an scalar field with a null kinetic term through the action:
\be
S=\frac{1}{2\kappa^2}\frac{1}{2\kappa^2}\int dx^4\sqrt{-g}\left(\phi R-V(\phi)\right)+S_m\ ,
\label{3.1}
\ee
Varying the action with respect to the scalar field, the corresponding equivalence is found:
\be
V'(\phi)=R\ \rightarrow\ \phi=\phi(R)\ ,\nn
f(R)=\phi(R)R-V(\phi(R)),
\label{3.2}
\ee
which yields the relations:
\be
\phi=f_R\ ,\quad V=Rf_R-f\ ,
\label{3.3}
\ee
As in the previous section, the reconstruction of the unimodular theory for the action (\ref{3.1}) is given by fixing the determinant of the metric,
\be
S=\frac{1}{2\kappa^2}\int dx^4\sqrt{-g}\left(\phi R-V(\phi)\right)-2\lambda(\sqrt{-g}-s_0)+S_m\ ,
\label{3.4}
\ee
The field equations are given by:
\be
R_{\mu\nu}-\frac{1}{2}g_{\mu\nu}\left(\phi R-V(\phi)\right)+\left(g_{\mu\nu}\Box-\nabla_{\mu}\nabla_{\nu}\right)\phi+g_{\mu\nu}\lambda=\kappa^2T_{\mu\nu}^{(m)}\ ,
\label{3.5}
\ee
Taking the divergence of the field equations, the condition $\nabla_{\mu}\lambda=0$ turns out and the integrability condition (\ref{2.7}) is obtained, which now is given by,
\be
\phi R-2V-3\Box\phi+\kappa^2T^{(m)}=4\lambda_0\ . 
\label{3.6}
\ee
Consequently, the field equations (\ref{3.5}) become the usual equations for the scalar-tensor theory (\ref{3.1}) with an additional cosmological constant. The question now arises, does the action (\ref{3.4}) have a counterpart in the Einstein frame? To do so, let us transform the action (\ref{3.4}) into the Einstein frame, what basically means recovering the usual Hilbert-Einstein action by applying the following conformal transformation:
\be
\tilde{g}_{\mu\nu}=\Omega^2g_{\mu\nu}\ \quad \text{where} \quad \Omega^2=\phi\ .
\label{3.7}
\ee
Here the tilde refers to the Einstein frame. Then, the Ricci scalar is transformed as follows:
\be
\tilde{R}=\frac{2}{\Omega^2}\left(R-\frac{6\Box\Omega}{\Omega}\right)\ .
\label{3.8}
\ee
And the action (\ref{3.4}) becomes
\begin{widetext}
\be
\tilde{S}=\int dx^4\left[\sqrt{-\tilde{g}}\left(\frac{\tilde{R}}{2\kappa^2}-\frac{1}{2}\partial_{\mu}\varphi\partial^{\mu}\varphi-\tilde{V}(\varphi)\right)-2\tilde{\lambda}\left(\sqrt{-\tilde{g}}e^{-2\sqrt{2/3}\kappa\varphi}-s_0\right)\ ,\right]
\label{3.9}
\ee
\end{widetext}
where we have redefined the scalar field,
\be
\phi=e^{\sqrt{2/3}\kappa\varphi}\ \quad \tilde{V}(\varphi)=\frac{e^{-2\sqrt{2/3}\kappa\varphi}}{2\kappa^2}V(\varphi)\ , \quad \tilde{\lambda}=\frac{\lambda}{2\kappa^2}\ .
\label{3.10}
\ee
The field equations are obtained by varying the action with respect to the metric,
\be
\tilde{R}_{\mu\nu}-\frac{1}{2}\tilde{g}_{\mu\nu}\tilde{R}=\kappa^2\left(T_{\mu\nu}^{(\varphi)}+T_{\mu\nu}^{(m)}\right)\ ,
\label{3.11}
\ee
where we have defined the energy-momentum tensor of the scalar field as,
\be
 T_{\mu\nu}^{(\varphi)}=\partial_{\mu}\varphi\partial_{\nu}\varphi-g_{\mu\nu}\left(\frac{1}{2}\partial_{\sigma}\varphi\partial^{\sigma}\varphi+\tilde{V}\right)-2\tilde{\lambda} g_{\mu\nu}e^{-2\sqrt{2/3}\kappa\varphi}
 \label{3.11b}
 \ee
The scalar field equation is obtained by varying the action (\ref{3.9}) with respect to the scalar field,
\be
\Box\varphi-V'(\varphi)+4\tilde{\lambda}\sqrt{\frac{2}{3}}\kappa e^{-2\sqrt{2/3}\kappa\varphi}=0\ .
\label{3.12}
\ee
While the variation of the action with respect to the Lagrange multiplier leads to the constraint,
\be
\sqrt{\tilde{-g}}=s_0\times e^{2\sqrt{2/3}\kappa\varphi}\ .
\label{3.13}
\ee
Hence, contrary to the case of the Jordan frame, the determinant of the metric $\tilde{g}_{\mu\nu}$ is not constant. Taking the divergence of the field equations (\ref{3.12}), and applying the identity $\nabla_{\mu}\left(\tilde{R}^{\mu\nu}-\frac{1}{2}\tilde{g}^{\mu\nu}\tilde{R}\right)=0$ and the matter-energy conservation $\nabla_{\mu}T^{\mu\nu(m)}=0$,  yields,
\bea
\nabla_{\mu}T^{\mu\nu(\varphi)}&=&\left(\Box\varphi-V'+4\tilde{\lambda}\sqrt{\frac{2}{3}}\kappa e^{-2\sqrt{2/3}\kappa\varphi}\right)\partial^{\nu}\varphi\nn
&&-2e^{-2\sqrt{2/3}\kappa\varphi}\partial^{\nu}\tilde{\lambda}=0\ .
\label{3.14}
\eea 
The first term in (\ref{3.14}) is the scalar field equation (\ref{3.12}), which becomes null leading to,
\be
\partial_{\nu}\tilde{\lambda}=0\ , \quad \rightarrow \tilde{\lambda}=\tilde{\lambda}_0\ .
\label{3.15}
\ee
Then, the energy-momentum tensor for the scalar field (\ref{3.11b}) turns out,
\be
T_{\mu\nu}^{(\varphi)}=\partial_{\mu}\varphi\partial_{\nu}\varphi-g_{\mu\nu}\left(\frac{1}{2}\partial_{\sigma}\varphi\partial^{\sigma}\varphi+\tilde{V}\right)-2\tilde{\lambda}_0 g_{\mu\nu}e^{-2\sqrt{2/3}\kappa\varphi}
\label{3.16}
\ee
Hence, the field equations (\ref{3.11}) are basically the equations of the action,
\be
\tilde{S}=\frac{1}{2\kappa^2}\int dx^4\left[\sqrt{-\tilde{g}}\left(\frac{\tilde{R}}{2\kappa^2}-\frac{1}{2}\partial_{\mu}\varphi\partial^{\mu}\varphi-\tilde{V}_{eff}(\varphi)\right)\right]\ ,
\label{3.17}
\ee
where the effective potential is defined as,
\be
\tilde{V}_{eff}(\varphi)=\tilde{V}(\varphi)+2\tilde{\lambda}_0 e^{-2\sqrt{2/3}\kappa\varphi}\ .
\label{3.18}
\ee
In comparison with the case in the Jordan frame, where a cosmological constant naturally arises, here the scalar potential is modified. what may introduce corrections to some solutions.

In the next section, we explore some cosmological solutions within the context of $f(R)$ and modified Gauss-Bonnet gravities, but also solutions in the Einstein frame are analyzed, particularly we study Starobinsky inflation within the context of unimodular gravity by applying the results obtained above.\\

%%%%%%%%%%%%%%%%%%%%%%%%%%%%%%%%%%%%%%%%%%%%%%%%%%%%%%
\section{Cosmological solutions}
\label{cosmo_sect}
%%%%%%%%%%%%%%%%%%%%%%%%%%%%%%%%%%%%%%%%%%%%%%%%%%%%%%
Let us now explore some cosmological solutions in the generalizations of unimodular gravity studied above. Here we intend to analyze dark energy solutions as well as some inflationary models.

%%%%%%%%%%%%%%%%%%%%%%%%%%%%%%%%%%%%%%%%%%%%%%%%%%%%%%
\subsection{Late-time acceleration}
%%%%%%%%%%%%%%%%%%%%%%%%%%%%%%%%%%%%%%%%%%%%%%%%%%%%%%
Since we are interested in late-time cosmological solutions, we assume a flat Friedmann-Lema\^itre-Robertson.Walker metric,
\be
ds^2=-dt^2+a^2(t)\sum_{i=1}^{3}dx^{i\;\;2}
\label{4.1}
\ee
Let us start by studying solutions in $f(R)$ unimodular gravity, whose field equations (\ref{2.8}) for the metric (\ref{4.1}) turns out
\begin{widetext}
\bea
H^2&=&\frac{1}{3f_R}\left[\kappa^2 \rho_m +\frac{Rf_R-f}{2}-3H\dot{R}f_{RR}+\lambda_0\right]\ , \nn
-3H^2-2\dot{H}&=&\frac{1}{f_R}\left[\kappa^2p_m+\dot{R}^2f_{RRR}+2H\dot{R}f_{RR}+\ddot{R}f_{RR}+\frac{1}{2}(f-Rf_R-2\lambda_0)\right]\ ,
\label{4.2}
\eea
\end{widetext}
Note that every solution of a particular $f(R)$ gravity is also a solution of unimodular $f(R)$ gravity just by shifting the action $f\rightarrow f+2\lambda_0$. Nevertheless, the additional constant in the FLRW equations (\ref{4.2}) may provide a wider set of solutions. In order to show so, let us analyze some particular and illustrative cosmological solutions. Since the universe goes through several accelerating stages, de Sitter solution plays an important role, where the Hubble parameter is given by
\be
H(t)=H_0
\label{4.3}
\ee
Moreover, $H=H_0$ is a critical point in every $f(R)$ gravity \cite{Cognola:2008zp}, such that the possible critical points of a particular gravitational action can be identified with the dark energy epoch and also inflation. Then, for the de Sitter solution (\ref{4.3}), the first FLRW (in vacuum) is given by,
\be
3f_{R_0}H_0^2-\frac{1}{2}(R_0f_{R_0}-f_0-\lambda_0)=0\ .
\label{4.4}
\ee
Hence, every root of this equation is a critical point and becomes a possible de Sitter stage along which the universe may go through. The presence of $\lambda_0$ introduces a correction that some particular $f(R)'s$, which lead to effective cosmological constant (as the Hu-Sawicky model \cite{Hu:2007nk}), may requires.\\

Let us now explore power-law solutions in cosmology, which also have a great importance along the universe history,
\be
a(t)=a_0t^m\ ,\quad H(t)=\frac{m}{t}\ ,
\label{4.5}
\ee
Note that for pressureless matter $m=2/3$, for radiation $m=1/2$ and for an accelerating universe $m>1$. The above solution has been analyzed in standard $f(R)$ gravity, where the following action holds \cite{Goheer:2009ss},
\be
f(R)=A_{\pm}R^{\frac{1}{4}\left(3-m\pm\sqrt{1+10m+m^2}\right)}\ .
\label{4.6}
\ee
Whether we assume the above $f(R)$ gravity in unimodular gravity with $m<1$, the effective cosmological constant $\lambda_0$ may become important at late-times, when the dark energy epoch starts, while the terms in (\ref{4.6}) may contribute during the matter/radiation epochs when they dominate over $\lambda_0$. Moreover, whether $m>1$ the unimodular $f(R)$ gravity (\ref{4.6}) contributes to the acceleration of the expansion, leading to corrections over a de Sitter expansion which would depend on the weight of $A_{\pm}$ in comparison with $\lambda_0$. \\

Let us now consider the unimodular version of Gauss-Bonnet gravity (\ref{2.11}), whose FLRW equation becomes:
\be
3H^2=\kappa^2\rho_m+\frac{1}{2}(Gf_G-f)-12f_{GG}\dot{G}H^3+\lambda_0\ ,
\label{4.7}
\ee
where $G=24(\dot{H}H^2+H^4)$. As in the previous case, we can analyze de Sitter solutions (\ref{4.3}) by introducing (\ref{4.3}) into the equation (\ref{4.7}), which turns out an algebraic equation,
\be
3H_0^2+\frac{1}{2}(f_0-G_0f_{G_0})-\lambda_0=0\ .
\label{4-8}
\ee
Hence, the merging cosmological constant $\lambda_0$ would determine the de Sitter points, and consequently the accelerating stages of the universe. In the case of power-law solutions (\ref{4.5}), the exact action within pure Gauss-Bonnet gravity (with no Ricci scalar in the action) that reproduces such solutions in vacuum are \cite{Myrzakulov:2010gt}:
\be
f(G)=AG^{\frac{1-m}{4}}\ ,
\label{4.9}
\ee
which may play the same role as in the case of $f(R)$ gravity, as shown above. Nevertheless, the most important feature of the action $f(R,G)=R+f(G)$ is that reproduces exact $\Lambda$CDM model,
\be
H^2=\frac{\Lambda}{3}+\frac{\kappa^2}{3}\rho_0 a^{-3}\ ,
\label{4.10}
\ee
by means of the gravitational action given by \cite{Elizalde:2010jx},
\bea
f(R,G)&=&R+a_1\left(\Lambda\pm\sqrt{9\Lambda^2-3G}\right)^2\nn
&&+a_2\left(\Lambda\pm\sqrt{9\Lambda^2-3G}\right)+a_3\ ,
\label{4.11}
\eea
where $a_{1}$ is an integration constant, $a_2=\frac{6-30a_1\Lambda}{15}$ and $a_3=3\left(1-6a_1\Lambda\right)$ are constants. Then, by identifying the last term of (\ref{4.11}) with the cosmological constant $\lambda_0$,
\be
\lambda_0=-\frac{3}{2}\left(1-6a_1\Lambda\right)
\label{4.12}
\ee
The unimodular version of Gauss-Bonnet gravity described by the action (\ref{4.11}) arises naturally as the gravitational action which leads to the $\Lambda$CDM model (\ref{4.10}). \\

Hence, extensions of unimodular gravity provide reliable descriptions of the late-time acceleration in a natural way.

%%%%%%%%%%%%%%%%%%%%%%%%%%%%%%%%%%%%%%%%
\subsection{Inflation}
%%%%%%%%%%%%%%%%%%%%%%%%%%%%%%%%%%%%%%%%

Let us now study how these extensions of unimodular gravity may affect the inflationary paradigm. In particular, here we analyze Starobinsky inflation \cite{staro} when considering the unimodular $f(R)$ theory (\ref{2.3}), which for the case of Starobinsky inflation is given by
\be
S=\frac{1}{2\kappa^2}\int dx^4\left[\sqrt{-g}\left( R+\frac{R^2}{6m^2}\right)-2\lambda(\sqrt{-g}-s_0)\right]\ ,
\label{4.13}
\ee
where $m^2$ is a constant. In order to simplify the calculations, we work in the scalar-tensor equivalence (\ref{3.4}), whose correspondence to the action (\ref{4.13}) is provided by
\be 
\phi=1+\frac{R}{3m^2}\ , \quad V(\phi)=3m^2(\phi-1)^2\ .
\label{4.14}
\ee
Applying the conformal transformation (\ref{3.7}) and the definitions (\ref{3.10}),  the action (\ref{3.17}) is constructed following the steps described in Section \ref{Einstein_sect},
%\be
%\tilde{S}=\int dx^4\left[\sqrt{-\tilde{g}}\left(\frac{\tilde{R}}{2\kappa^2}-\frac{1}{2}\partial_{\mu}\varphi\partial^{\mu}\varphi-\tilde{V}_{eff}(\varphi)\right)\right]\ ,
%\label{4.15}
%\ee
where the effective potential for the case (\ref{4.13}) is given by,
\be
\tilde{V}_{eff}(\varphi)=\frac{1}{2\kappa^2}\left[\frac{3m^2}{2}\left(1-e^{-2\sqrt{2/3}\kappa\varphi}\right)^2-2\lambda_0 e^{-2\sqrt{2/3}\kappa\varphi}\right]\ .
\label{4.16}
\ee
Then, the FLRW equations are:
\bea
\frac{3}{\kappa^2} H^2&=& \frac{1}{2}{\dot \varphi}^2 + \tilde{V}_{eff}(\varphi)\, ,  \nn 
- \frac{1}{\kappa^2} \left( 3 H^2 + 2\dot H \right) &=& \frac{1}{2}{\dot \varphi}^2 - \tilde{V}_{eff}(\varphi)\, ,
\label{4.17}
\eea 
While the scalar field satisfies
\be
\ddot{\varphi}+3H\dot{\varphi}+\frac{\partial \tilde{V}_{eff}(\varphi)}{\partial\varphi}=0
\label{4.18}
\ee
Slow-roll inflation occurs in the regime $\kappa \varphi \gg 1$, where the friction term in (\ref{4.18}) dominates, and the expansion grows exponentially approximately, being the Hubble parameter $H\sim H_0$. Then, the following relations hold   
\be
H\dot{\phi}\gg\ddot{\varphi}\,,\quad \tilde{V}\gg\dot{\varphi}^2\ .
\label{4.19}
\ee
Equivalently, we can define the slow-roll parameters,
\be
\epsilon=
\frac{1}{2\kappa^2} \left( \frac{\tilde{V}_{eff}'(\varphi)}{\tilde{V}_{eff}(\varphi)} \right)^2\, ,\quad 
\eta= \frac{1}{\kappa^2} \frac{\tilde{V}_{eff}''(\varphi)}{\tilde{V}_{eff}(\varphi)}\, , \quad 
\label{4.19b} 
\ee
Hence, during inflation $\epsilon\ll 1$ and $\eta<1$, while after an enough number of e-foldings, usually around $N=50-65$, $\epsilon\geq 1$, when the scalar field $\varphi$ rolls down the potential slope and the kinetic term becomes important and eventually dominates. Then, the field oscillates around the minimum of the potential, emitting particles and reheating the Universe. Hence, by using these approximations and combining the FLRW equations (\ref{4.17}) and the scalar field equation (\ref{4.18}), the equations during inflation are given approximately by,
\bea
 H^2  \simeq  \frac{\kappa^2}{3}  \tilde{V}_{eff}(\varphi), \nn
 3H \dot{\varphi} \simeq -  \tilde{V}_{eff}'(\varphi)
 \label{4.20}
 \eea
The slow-roll parameters (\ref{4.19b}) for the potential (\ref{4.16}) are given by,
\bea
\epsilon&=&\frac{4}{3} \frac{\left[3m^2\left(-1+e^{\sqrt{2/3}\kappa\varphi}\right)-4\lambda_0\right]^2}{\left[3m^2\left(-1+e^{\sqrt{2/3}\kappa\varphi}\right)^2+4\lambda_0\right]^2}\ , \\ \nonumber
\eta&=&\frac{4}{3} \frac{-3m^2\left(-2+e^{\sqrt{2/3}\kappa\varphi}\right)+8\lambda_0}{3m^2\left(-1+e^{\sqrt{2/3}\kappa\varphi}\right)^2+4\lambda_0}\ ,
\label{4.21}
\eea
Starobinsky inflation is recovered by setting $\lambda_0=0$. Nevertheless, since $m^2/\lambda_0>\e^{-2\sqrt{2/3}\kappa\varphi_{start}}$ in order to ensure an enough number of e-foldings before the field rolls down, together with $\kappa\varphi\gg 1$, it gives the following the slow-roll parameters,
\bea
\epsilon&=&\frac{4}{3} e^{-2\sqrt{2/3}\kappa\varphi}\ , \\ \nonumber
\eta&=&-\frac{4}{3} e^{-\sqrt{2/3}\kappa\varphi}\ ,
\label{4.22}
\eea
In addition, the spectral index and the scalar-to-tensor ratio are given in terms of the slow-roll parameters by
\be
n_s-1=-3\epsilon+2\eta\ \quad r=16\epsilon\ .
\label{4.23}
\ee
It is straightforward to calculate the number of e-foldings during inflation, which is given by
\be
N  \equiv  \int_{t_{start}}^{t_{end}} \tilde{H} dt=-\kappa^2 \int_{\varphi_{start}}^{\varphi_{end}} \frac{\tilde{V}_{eff}(\varphi)}{\tilde{V}_{eff}'(\varphi)} 
\simeq \frac{3}{4}e^{\sqrt{2/3}\kappa \; \tilde{\varphi}_{start}}.
\label{4.24}
\ee
Note that the number of e-foldings is related to the slow-roll parameters as 
\be
\epsilon \simeq \frac{3}{4}\frac{1}{N^2}\ , \quad \eta \simeq -\frac{1}{N}\ .
\label{SRparamStaro}
\ee
Then, assuming a number of e-foldings $N\sim65$,the following values of the inflationary observables are obtained:
%%%%%%%%
\be
n_s=0.968\ , \quad r=0.00284\ .
\label{SpRStaro}
\ee

This is exactly the result as in Starobisnky inflation, which satisfies quite well the constraints provided by the last data \cite{Ade:2015lrj}. Hence, as far as $m^2/\lambda_0>\e^{-2\sqrt{2/3}\kappa\varphi_{start}}$, the unimodular version of Starobinsky inflation is likewise successful, but also includes the corresponding cosmological constant that may dominate at late-times, leading to a complete description of the universe evolution.

%%%%%%%%%%%%%%%%%%%%%%%%%%%%%%%%%%%%%%%%%%%%%%%%%%%%%%
\section{Conclusions}
\label{Conclusions_sect}
%%%%%%%%%%%%%%%%%%%%%%%%%%%%%%%%%%%%%%%%%%%%%%%%%%%%%%

Summarizing, in this manuscript we have extended the so-called unimodular gravity to other more general actions rather than the Hilbert-Einstein action. As in the original case, extensions of unimodular gravity can be constructed departing from the trace-free part of the field equations or alternatively by the gauge choice that fixes the determinant of the metric to be a constant. As widely pointed in the literature, any implementation of the unimodular constraint lead to the same results at the classical level, but may provide differences when quantum mechanics are considered. Nevertheless, since the paper is devoted to classical aspects, we have forced the unimodular constraint in the action through a Lagrange multiplier, such that calculations become simpler when dealing gravitational Lagrangians with more general functions of curvature invariants. Hence, following this procedure, and in spite of the apparent lack of symmetries, extensions of unimodular gravity lead to the covariant field equations of the originals with the presence of a cosmological constant. \\

The issue is more subtle when dealing with conformal transformations. As shown, by transforming the gravitational action from the Jordan to the Einstein frame, the determinant is not fixed to be a constant anymore. However, the Lagrange multiplier used to fix the determinant of the metric, turns out a constant as well, such that the corresponding counterpart in the Einstein frame becomes the usual  quintessence-like model but in this case with a correction in the scalar potential. Such additional term may have consequence when studying some particular solutions.\\

Finally, some cosmological solutions have been studied within the unimodular version of Gauss-Bonnet gravity and $f(R)$ gravity (together with its scalar-tensor equivalence). As shown, unimodular version of these theories provides a richer set of solution and is able to give a complete picture of the universe evolution in a natural way. In addition, predictions from Starobinsky inflation are fully recovered as far as the correction in the scalar potential is well set. Moreover, the unimodular version of Starobinsky inflation may provide an explanation to the late-time acceleration through the effective cosmological constant that naturally arises. Hence, such results point to $R+R^2$ as a reliable cosmological model for describing the whole universe history.

%%%%%%%%%%%%%%%%%%%%%%%%
%%%  Acknowledgments
%%%%%%%%%%%%%%%%%%%%%%%%
\section*{Acknowledgments}
I would like to thank Sergei D. Odintsov and Ippocratis Saltas for his help and valuable comments. I acknowledge support from a postdoctoral fellowship Ref.~SFRH/BPD/95939/2013 by Funda\c{c}\~ao para a Ci\^encia e a Tecnologia (FCT, Portugal) and the support through the research grant UID/FIS/04434/2013 (FCT, Portugal).

%%%%%%%%%%%

%%%%%%%%%%%

% \bibliographystyle{PRD}


\begin{thebibliography}{99}

%Cosmological constant problem


\bibitem{Weinberg:1988cp} 
  S.~Weinberg,
  %``The Cosmological Constant Problem,''
  Rev.\ Mod.\ Phys.\  {\bf 61}, 1 (1989).
 % doi:10.1103/RevModPhys.61.1
  %%CITATION = doi:10.1103/RevModPhys.61.1;%%
  %3096 citations counted in INSPIRE as of 03 Feb 2016

\bibitem{Padilla:2015aaa} 
  A.~Padilla,
  %``Lectures on the Cosmological Constant Problem,''
  arXiv:1502.05296 [hep-th].
  %%CITATION = ARXIV:1502.05296;%%
%\bibitem{Peebles:2002gy} 
  P.~J.~E.~Peebles and B.~Ratra,
  %``The Cosmological constant and dark energy,''
  Rev.\ Mod.\ Phys.\  {\bf 75}, 559 (2003)
  %doi:10.1103/RevModPhys.75.559
  [astro-ph/0207347];
  %\bibitem{Burgess:2013ara} 
  C.~P.~Burgess,
  %``The Cosmological Constant Problem: Why it's hard to get Dark Energy from Micro-physics,''
  %doi:10.1093/acprof:oso/9780198728856.003.0004
  arXiv:1309.4133 [hep-th].


%Dark energy reviews

\bibitem{dark_energy}
E.~J.~Copeland, M.~Sami and S.~Tsujikawa,
  %``Dynamics of dark energy,''
  Int.\ J.\ Mod.\ Phys.\ D {\bf 15}, 1753 (2006)
  %doi:10.1142/S021827180600942X
  [hep-th/0603057]:
  K. Bamba, S. Capozziello, S. Nojiri and S. D. Odintsov, Astrophys. Space Sci. 342, 155 (2012) [arXiv:1205.3421 [gr-qc]]
  L.~Amendola, \textit{Dark Energy: Theory and Observations}, Cambrigde Press 2015;
  Phys.\ Rept.\  {\bf 509}, 167 (2011)   [arXiv:1108.6266 [gr-qc]];
  %
%  \bibitem{Peebles:2002gy}
  F.~S.~N.~Lobo, %``The Dark side of gravity: Modified theories of gravity,''
  Dark Energy-Current Advances and Ideas
  [arXiv:0807.1640 [gr-qc]];
  %%CITATION = ARXIV:0807.1640;%%
%\cite{Nojiri:2010wj}
   %%CITATION = ARXIV:1402.7114;%%

\bibitem{Nojiri:2010wj} 
  S.~Nojiri and S.~D.~Odintsov,
%  ``Unified cosmic history in modified gravity: from F(R) theory to Lorentz non-invariant  models,''
  Phys.\ Rept.\  {\bf 505}, 59 (2011)   [arXiv:1011.0544 [gr-qc]]; eConf C {\bf 0602061} (2006) 06   [Int.\ J.\ Geom.\ Meth.\ Mod.\ Phys.\ {\bf 4} (2007) 115] [hep-th/0601213]; 
T.~P.~Sotiriou and V.~Faraoni, %`f(R) Theories Of Gravity,''
  Rev.\ Mod.\ Phys.\  {\bf 82}, 451 (2010)   [arXiv:0805.1726 [gr-qc]];
%\bibitem{DeFelice:2010aj} 
  A.~De Felice and S.~Tsujikawa, %   ``f(R) theories,''
  Living Rev.\ Rel.\  {\bf 13}, 3 (2010)   [arXiv:1002.4928 [gr-qc]];
  %
  S.~Capozziello and M.~De Laurentis, %``Extended Theories of Gravity,''
  Phys.\ Rept.\  {\bf 509}, 167 (2011)   [arXiv:1108.6266 [gr-qc]];
  %
%\bibitem{Bamba:2014eea}
   K.~Bamba and S.~D.~Odintsov,  %``Universe acceleration in modified gravities:
   %$F(R)$ and $F(T)$ cases,''
   arXiv:1402.7114 [hep-th]; 
   %%CITATION = ARXIV:1402.7114;%%
%\bibitem{Book-Capozziello-Faraoni}
S.~Capozziello and V.~Faraoni, \textit{Beyond Einstein Gravity}, Springer, Dordrecht, (2010) ;
%%CITATION = ARXIV:1108.6266;%%
%\bibitem{delaCruzDombriz:2012xy} 
%\bibitem{SaezGomez:2011yp} 
  D.~Saez-Gomez,
  %``Scalar-tensor theory with Lagrange multipliers: a way of understanding the cosmological constant problem, and future singularities,''
  Phys.\ Rev.\ D {\bf 85}, 023009 (2012)
  %doi:10.1103/PhysRevD.85.023009
  [arXiv:1110.6033 [hep-th]];
  A.~de la Cruz-Dombriz and D.~Saez-Gomez,
%  ``Black holes, cosmological solutions, future singularities, and their thermodynamical properties in modified gravity theories,''
  Entropy {\bf 14}, 1717 (2012) [arXiv:1207.2663 [gr-qc]].



%Unimodular gravity

\bibitem{Finkelstein:2000pg} 
  D.~R.~Finkelstein, A.~A.~Galiautdinov and J.~E.~Baugh,
  %``Unimodular relativity and cosmological constant,''
  J.\ Math.\ Phys.\  {\bf 42}, 340 (2001)
  %doi:10.1063/1.1328077
  [gr-qc/0009099];
   % \bibitem{Ellis:2010uc} 
  G.~F.~R.~Ellis, H.~van Elst, J.~Murugan and J.~P.~Uzan,
  %``On the Trace-Free Einstein Equations as a Viable Alternative to General Relativity,''
  Class.\ Quant.\ Grav.\  {\bf 28}, 225007 (2011)
  %doi:10.1088/0264-9381/28/22/225007
  [arXiv:1008.1196 [gr-qc]];
%\bibitem{Kluson:2014esa} 
  J.~Kluson,
  %``Canonical Analysis of Unimodular Gravity,''
  Phys.\ Rev.\ D {\bf 91}, no. 6, 064058 (2015)
  %doi:10.1103/PhysRevD.91.064058
  [arXiv:1409.8014 [hep-th]];
    %\bibitem{Barcelo:2014mua} 
  C.~Barcelo, R.~Carballo-Rubio and L.~J.~Garay,
  %``Unimodular gravity and general relativity from graviton self-interactions,''
  Phys.\ Rev.\ D {\bf 89}, no. 12, 124019 (2014)
%  doi:10.1103/PhysRevD.89.124019
  [arXiv:1401.2941 [gr-qc]];
  %\bibitem{Burger:2015kie} 
  D.~J.~Burger, G.~F.~R.~Ellis, J.~Murugan and A.~Weltman,
  %``The KLT relations in unimodular gravity,''
  arXiv:1511.08517 [hep-th];
%  \bibitem{Gao:2014nia} 
  C.~Gao, R.~H.~Brandenberger, Y.~Cai and P.~Chen,
  %``Cosmological Perturbations in Unimodular Gravity,''
  JCAP {\bf 1409}, 021 (2014)
 % doi:10.1088/1475-7516/2014/09/021
  [arXiv:1405.1644 [gr-qc]].
  %\bibitem{Cho:2014taa} 
  I.~Cho and N.~K.~Singh,
  %``Unimodular Theory of Gravity and Inflation,''
  Class.\ Quant.\ Grav.\  {\bf 32}, no. 13, 135020 (2015)
  %doi:10.1088/0264-9381/32/13/135020
  [arXiv:1412.6205 [gr-qc]].
 % \bibitem{Jain:2012gc} 
  P.~Jain, A.~Jaiswal, P.~Karmakar, G.~Kashyap and N.~K.~Singh,
  %``Cosmological implications of unimodular gravity,''
  JCAP {\bf 1211}, 003 (2012)
  %doi:10.1088/1475-7516/2012/11/003
  [arXiv:1109.0169 [astro-ph.CO]];

\bibitem{Alvarez:2005iy} 
  E.~Alvarez,
  %``Can one tell Einstein's unimodular theory from Einstein's general relativity?,''
  JHEP {\bf 0503}, 002 (2005)
  %doi:10.1088/1126-6708/2005/03/002
  [hep-th/0501146].  

\bibitem{Padilla:2014yea} 
%\bibitem{Smolin:2010iq} 
  L.~Smolin,
  %``Unimodular loop quantum gravity and the problems of time,''
  Phys.\ Rev.\ D {\bf 84}, 044047 (2011)
 % doi:10.1103/PhysRevD.84.044047
  [arXiv:1008.1759 [hep-th]];
%\bibitem{Smolin:2009ti} 
 % L.~Smolin,
  %``The Quantization of unimodular gravity and the cosmological constant problems,''
  Phys.\ Rev.\ D {\bf 80}, 084003 (2009)
  %doi:10.1103/PhysRevD.80.084003
  [arXiv:0904.4841 [hep-th]];
%\bibitem{Unruh:1988in} 
  W.~G.~Unruh,
  %``A Unimodular Theory of Canonical Quantum Gravity,''
  Phys.\ Rev.\ D {\bf 40}, 1048 (1989);
  %doi:10.1103/PhysRevD.40.1048
% [12]
% W. G. Unruh and R. M. Wald, “Time and the Interpreta-
% tion of Canonical Quantum Gravity,”
% Phys.Rev.
% , vol. D40,
% p. 2598, 1989.
% [13]
% L. Bombelli, W. Couch, and R. Torrence, “Time as space-
% time four volume and the Ashtekar variables,”
% Phys.Rev.
% ,
% vol. D44, pp. 2589–2592, 1991.
%\bibitem{Kuchar:1991xd} 
  K.~V.~Kuchar,
  %``Does an unspecified cosmological constant solve the problem of time in quantum gravity?,''
  Phys.\ Rev.\ D {\bf 43}, 3332 (1991);
  %doi:10.1103/PhysRevD.43.3332
%\bibitem{Alvarez:2012px} 
  E.~Alvarez and M.~Herrero-Valea,
  %``Unimodular gravity with external sources,''
  JCAP {\bf 1301}, 014 (2013)
  %doi:10.1088/1475-7516/2013/01/014
  [arXiv:1209.6223 [hep-th]];
 %\bibitem{Alvarez:2013fs} 
  %E.~Alvarez and M.~Herrero-Valea,
  %``No Conformal Anomaly in Unimodular Gravity,''
  Phys.\ Rev.\ D {\bf 87}, 084054 (2013)
  %doi:10.1103/PhysRevD.87.084054
  [arXiv:1301.5130 [hep-th]]; 
  
%\bibitem{Alvarez:2015pla} 
  E.~Alvarez, S.~Gonz\'alez-Mart\'in, M.~Herrero-Valea and C.~P.~Mart\'in,
  %``Unimodular Gravity Redux,''
  Phys.\ Rev.\ D {\bf 92}, no. 6, 061502 (2015)
 % doi:10.1103/PhysRevD.92.061502
  [arXiv:1505.00022 [hep-th]];
  %\bibitem{Alvarez:2015sba} 
 % E.~Álvarez, S.~González-Martín, M.~Herrero-Valea and C.~P.~Martín,
  %``Quantum Corrections to Unimodular Gravity,''
  JHEP {\bf 1508}, 078 (2015)
  %doi:10.1007/JHEP08(2015)078
  [arXiv:1505.01995 [hep-th]];  
 B.~Fiol and J.~Garriga,
  %``Semiclassical Unimodular Gravity,''
  JCAP {\bf 1008}, 015 (2010)
  %doi:10.1088/1475-7516/2010/08/015
  [arXiv:0809.1371 [hep-th]];
%\bibitem{Eichhorn:2013xr} 
  A.~Eichhorn,
  %``On unimodular quantum gravity,''
  Class.\ Quant.\ Grav.\  {\bf 30}, 115016 (2013)
  %doi:10.1088/0264-9381/30/11/115016
  [arXiv:1301.0879 [gr-qc]];
A.~Padilla and I.~D.~Saltas,
  %``A note on classical and quantum unimodular gravity,''
  Eur.\ Phys.\ J.\ C {\bf 75}, no. 11, 561 (2015)
  %doi:10.1140/epjc/s10052-015-3767-0
  [arXiv:1409.3573 [gr-qc]];
%\bibitem{Saltas:2014cta} 
  I.~D.~Saltas,
  %``UV structure of quantum unimodular gravity,''
  Phys.\ Rev.\ D {\bf 90}, no. 12, 124052 (2014)
  %doi:10.1103/PhysRevD.90.124052
  [arXiv:1410.6163 [hep-th]];
%\bibitem{Benedetti:2015zsw} 
  D.~Benedetti,
  %``Essential nature of Newton's constant in unimodular gravity,''
  arXiv:1511.06560 [hep-th].
 
  \bibitem{Alvarez:2007nn} 
  E.~Alvarez and A.~F.~Faedo,
  %``Unimodular cosmology and the weight of energy,''
  Phys.\ Rev.\ D {\bf 76}, 064013 (2007)
  %doi:10.1103/PhysRevD.76.064013
  [hep-th/0702184]. 
 
 
\bibitem{Alvarez:2009ga} 
  E.~Alvarez, A.~F.~Faedo and J.~J.~Lopez-Villarejo,
  %``Transverse gravity versus observations,''
  JCAP {\bf 0907}, 002 (2009)
  doi:10.1088/1475-7516/2009/07/002
  [arXiv:0904.3298 [hep-th]].    

\bibitem{Alvarez:2006uu} 
  E.~Alvarez, D.~Blas, J.~Garriga and E.~Verdaguer,
  %``Transverse Fierz-Pauli symmetry,''
  Nucl.\ Phys.\ B {\bf 756}, 148 (2006)
  doi:10.1016/j.nuclphysb.2006.08.003
  [hep-th/0606019].
  
  \bibitem{Alvarez:2008zw} 
  E.~Alvarez, A.~F.~Faedo and J.~J.~Lopez-Villarejo,
  %``Ultraviolet behavior of transverse gravity,''
  JHEP {\bf 0810}, 023 (2008)
  doi:10.1088/1126-6708/2008/10/023
  [arXiv:0807.1293 [hep-th]].
%\bibitem{Alvarez:2010cg} 
  E.~Alvarez and R.~Vidal,
  %``Weyl transverse gravity (WTDiff) and the cosmological constant,''
  Phys.\ Rev.\ D {\bf 81}, 084057 (2010)
  doi:10.1103/PhysRevD.81.084057
  [arXiv:1001.4458 [hep-th]].
 
\bibitem{Buchmuller:1988wx} 
  W.~Buchmuller and N.~Dragon,
  %``Einstein Gravity From Restricted Coordinate Invariance,''
  Phys.\ Lett.\ B {\bf 207}, 292 (1988).
  %doi:10.1016/0370-2693(88)90577-1
  %%CITATION = doi:10.1016/0370-2693(88)90577-1;%%

\bibitem{Henneaux:1989zc} 
  M.~Henneaux and C.~Teitelboim,
  %``The Cosmological Constant and General Covariance,''
  Phys.\ Lett.\ B {\bf 222}, 195 (1989).
  doi:10.1016/0370-2693(89)91251-3  

  
  
%Modified unimodular gravity

  \bibitem{Eichhorn:2015bna} 
  A.~Eichhorn,
  %``The Renormalization Group flow of unimodular f(R) gravity,''
  JHEP {\bf 1504}, 096 (2015)
  %doi:10.1007/JHEP04(2015)096
  [arXiv:1501.05848 [gr-qc]];
    


\bibitem{Nojiri:2015sfd} 
  S.~Nojiri, S.~D.~Odintsov and V.~K.~Oikonomou,
  %``Unimodular $F(R)$ Gravity,''
  arXiv:1512.07223 [gr-qc].
  %%CITATION = ARXIV:1512.07223;%%
%\bibitem{Nojiri:2016ygo} 
 % S.~Nojiri, S.~D.~Odintsov and V.~K.~Oikonomou,
  %``The bounce universe history from unimodular $F(R)$ gravity,''
  arXiv:1601.04112 [gr-qc].
  %%CITATION = ARXIV:1601.04112;%%
%\bibitem{Nojiri:2016ppu} 
  %S.~Nojiri, S.~D.~Odintsov and V.~K.~Oikonomou,
  %``Unimodular-Mimetic Cosmology,''
  arXiv:1601.07057 [gr-qc].

   \bibitem{staro}
Starobinsky, A.~A.
  Phys.\ Lett.\ B {\bf 91}, 99 (1980).

  \bibitem{Hu:2007nk} 
  W.~Hu and I.~Sawicki,
  %``Models of f(R) Cosmic Acceleration that Evade Solar-System Tests,''
  Phys.\ Rev.\ D {\bf 76}, 064004 (2007)
  %doi:10.1103/PhysRevD.76.064004
  [arXiv:0705.1158 [astro-ph]].
    
\bibitem{Elizalde:2010jx} 
  E.~Elizalde, R.~Myrzakulov, V.~V.~Obukhov and D.~Saez-Gomez,
  %``LambdaCDM epoch reconstruction from F(R,G) and modified Gauss-Bonnet gravities,''
  Class.\ Quant.\ Grav.\  {\bf 27}, 095007 (2010)
 % doi:10.1088/0264-9381/27/9/095007
  [arXiv:1001.3636 [gr-qc]].


\bibitem{Cognola:2008zp} 
  G.~Cognola, E.~Elizalde, S.~D.~Odintsov, P.~Tretyakov and S.~Zerbini,
  %``Initial and final de Sitter universes from modified f(R) gravity,''
  Phys.\ Rev.\ D {\bf 79}, 044001 (2009)
  %doi:10.1103/PhysRevD.79.044001
  [arXiv:0810.4989 [gr-qc]].

  
 \bibitem{Goheer:2009ss} 
  N.~Goheer, J.~Larena and P.~K.~S.~Dunsby,
  %``Power-law cosmic expansion in f(R) gravity models,''
  Phys.\ Rev.\ D {\bf 80}, 061301 (2009)
  %doi:10.1103/PhysRevD.80.061301
  [arXiv:0906.3860 [gr-qc]]. 

 \bibitem{Myrzakulov:2010gt} 
  R.~Myrzakulov, D.~Saez-Gomez and A.~Tureanu,
  %``On the $\Lambda$CDM Universe in $f(G)$ gravity,''
  Gen.\ Rel.\ Grav.\  {\bf 43}, 1671 (2011)
  %doi:10.1007/s10714-011-1149-y
  [arXiv:1009.0902 [gr-qc]].
  %%CITATION = doi:10.1007/s10714-011-1149-y;%%
 
  

\bibitem{Ade:2015lrj} 
  P.~A.~R.~Ade {\it et al.} [Planck Collaboration],
  %``Planck 2015 results. XX. Constraints on inflation,''
  arXiv:1502.02114 [astro-ph.CO].
%\bibitem{Array:2015xqh} 
  P.~A.~R.~Ade {\it et al.} [BICEP2 and Keck Array Collaborations],
  %``Improved Constraints on Cosmology and Foregrounds from BICEP2 and Keck Array Cosmic Microwave Background Data with Inclusion of 95 GHz Band,''
  Phys.\ Rev.\ Lett.\  {\bf 116}, no. 3, 031302 (2016)
  %doi:10.1103/PhysRevLett.116.031302
  [arXiv:1510.09217 [astro-ph.CO]].

\end{thebibliography}
\end{document}